\input{aipcheck}

\documentclass[
    ,final            
  ]
  {aipproc}

\layoutstyle{6x9}

\begin{document}

\title{Time-Delayed transfer functions simulations for LMXBs}


\author{T. Muñoz-Darias}{
  address={Instituto de Astrofísica de Canarias, 38200 La Laguna, Tenerife, Spain ; tmd@iac.es}
}

\author{I. G. Martínez-Pais}{
  address={Instituto de Astrofísica de Canarias, 38200 La Laguna, Tenerife, Spain}}

\author{J. Casares}{
  address={Instituto de Astrofísica de Canarias, 38200 La Laguna, Tenerife, Spain}}

\begin{abstract}
Recent works (Steeghs \& Casares 2002, Casares et al. 2003, Hynes et al. 2003) have demonstrated that Bowen flourescence is a very efficient tracer of the companion star in LMXBs. We present a numerical code to simulate time-delayed transfer functions in LMXBs, specific to the case of reprocessing in emission lines. The code is also able to obtain geometrical and binary parameters by fitting observed (X-ray + optical) light curves using simulated annealing methods. In this work we present the geometrical model for the companion star and the analytical model for the disc and show synthetic time-delay transfer functions for different orbital phases and system parameters.
\end{abstract}

\maketitle


\section{Introduction}

Optical emission in low mass X-ray binaries (LMXBs, hereafter) mainly arises from X-ray reprocessing in different binary sites. Therefore, the observed optical variability should be delayed due to light travel time difference, with respect to the source  X-ray emission.
Echo tomography is a very powerful technique which allows one to probe the accretion geometry in active LMXBs by studying time delay as a function of the orbital phase. Previous attempts using broad band photometry and spectroscopy (Hynes et al. 2003 , O`Brien 2001) have shown that the accretion disc is the dominant reprocessing site in the continuum.
On the other hand, narrow emission lines originating from the irradiated donor star have been discovered in Sco X-1 (Steeghs \& Casares 2002) and others LMXBs (Casares et al. 2004). These narrow features are strongest in the Bowen blend, a set of resonant emission lines, mainly CIII and NIII $\lambda$4640 \AA. In particular, the NIII line are powered by flourescence resonance, which require seed photons of HeII Ly$\alpha$. The Roche lobe shaped donor star intercepts the energetic photons from the inner accretion disc resulting in the observed optical emission lines from its surface. Therefore, these features are expected to show delayed variability with respect to the irradiating X-ray emission, which can be used to constrain the size of the binary. Since it is clear that Bowen blend emission arises from the companion star, we have built time-delayed transfer functions specific for reprocessing in spectral lines. These functions are based on a geometrical model for the companion star and analytic solutions for  the accretion disc.
\section{Geometrical Model for the Companion Star}
The model developed is based on the Roche potential for the secondary star. In order to carry out the numerical integration we have divided the shaped surface in small triangular tiles of equal area which cover completely the Roche lobe filling star (see figure 1). The mass ratio q $(M_2/M_1)$ is the free parameter which defines the shape of the companion. On the other hand, the inclination and the orbital phase determine in which moment the different regions are visible by the observer.
Eclipses and shadowing effects by a cilindrical accretion disc are also included in the model.
\begin{figure}
  \includegraphics[height=.3\textheight]{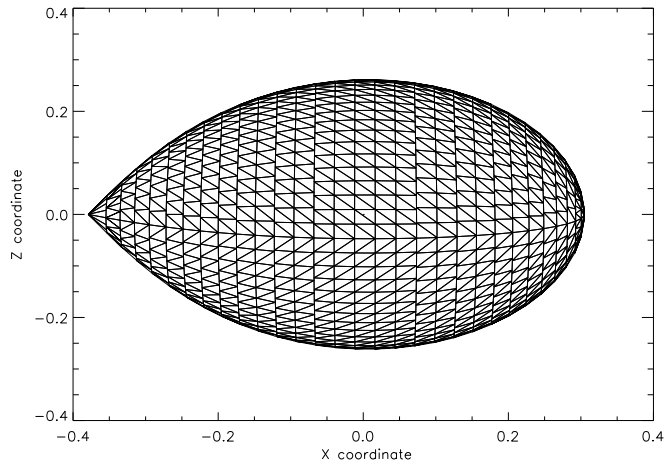}
  \caption{Roche lobe filling star for a mass ratio q = 0.3  and orbital phase of 0.25}
\end{figure}
\section{Building Time-Delayed Transfer Functions}
We have considered time-delayed transfer functions as the response given by the system to a normalized X-ray  emission. As reprocessing time scales for  spectral lines are much lower than 1 second the system response can be aproximated by a mirror function whose efficiency only depends on the albedo  and the X, Y, Z coordinates of the region considered.
For the case of the companion star it is necessary to define a delay-bin (given by the size of the tiles) and to sum up the contributions of the  irradiated regions which are visible by the observer.  This is repeated  for all the possible delays, resulting in time-delayed transfer functions for the companion star which depends on  the orbital phase, the inclination, the mass ratio q and the disc geometry.
In order to compute the Transfer Function for the disc, we have approximated  this by an axially symmetric structure whose vertical cilindrical coordinate, z, depends on the radial coordinate as $r^{\beta}$. The geometry of the disc is characterized by $\beta$ and both, the inner and outer disc radii. In this way, the transfer function can be obtained analytically taking into account the effects of albedo and the local projection factors. The exponent $\beta$ is fixed to 9/7, for the case of a standard irradiated disc (Vrtilik et al, 1990).\\

\begin{figure}
  \includegraphics[height=.35\textheight]{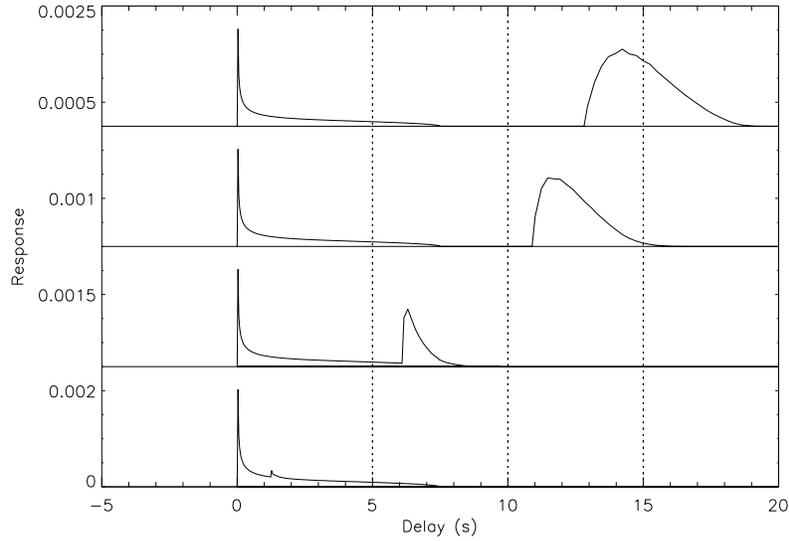}
  \caption{Time-delayed transfer functions for i = 80 and system parameters of Sco X-1 (i.e. q= 0.3 and P= 18.9 h). For phases close to 0.5 the delay is maximum (15 s) and the response of the companion star is comparable to the accretion disc contribution.}
\end{figure}
\begin{figure}
  \includegraphics[height=.35\textheight]{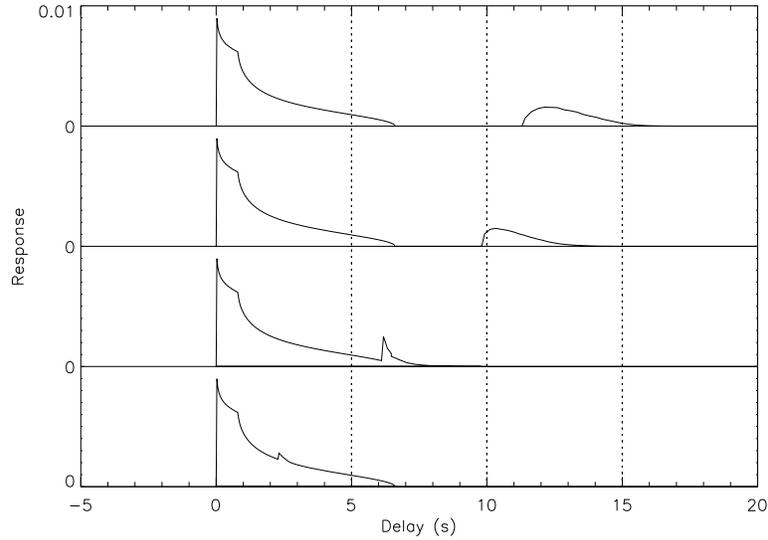}
  \caption{Same as in figure 2 but for i = 50. For lower inclinations the disc contribution dominates the transfer function. In this case, for phases close to 0.5, the maximun delay is 12.5 s which is lower than for i = 80 because it scales with cos (i).}
\end{figure}

After combining the disc and the companion star contributions we get the time-delayed transfer function for the entire system. As we see in figures 2, 3 and 4 the function is  strongly dependent on  both the orbital phase and system inclination. Therefore, it is possible to compare these simulations with real data (simultaneous X-ray/optical narrow band photometry) in order to constrain  the system parametres.


\begin{figure}
  \includegraphics[height=.35\textheight]{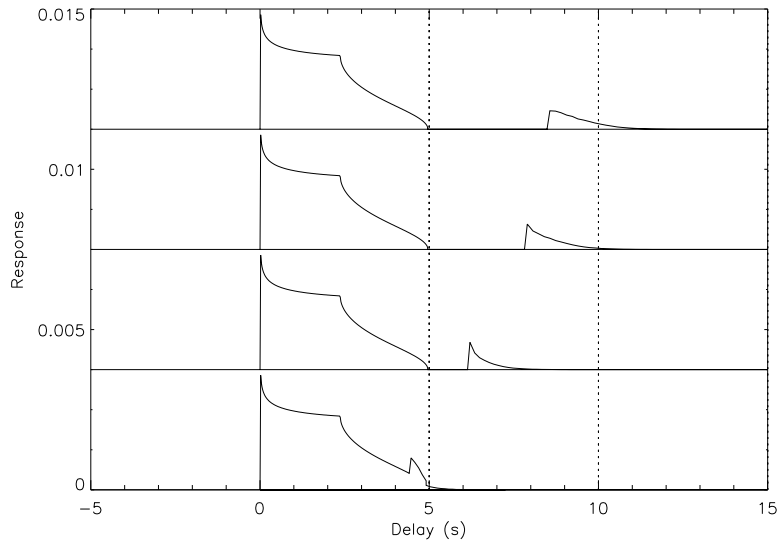}
  \caption{Same as in figure 2 but  for i=20.  The profile of the disc transfer functions is remarkably different in this case because the inner disc is fully visible.}
\end{figure}


\section{CONCLUSIONS}
\begin{enumerate}
\item
  As expected, our Time-delayed transfer functions are strongly dependent on the inclination, the mass ratio (q) and the disc parameters (radius and $\beta$ coefficient)
\item
 The transfer function is dominated by the disc contribution for i $\leq$ 60
\item
The profile of the disc transfer function is very sensitive to $\beta$ for very low inclinations
\end{enumerate}
\vspace{0.5cm}

\section{REFERENCES}
\footnotesize{ \hspace{4mm}Casares, J., Steeghs, D., Hynes, R. I., \& Charles, P. A. 2003, ApJ, 590, 1041
\smallskip

Casares, J., Steeghs, D., Hynes, R. I., Charles, P. A., Cornelisse R., \& O'Brien, K. 2004, Rev.Mex AA,30,21
\smallskip

Hynes, R. I., Steeghs D., Casares, J., Charles, P. A., \& O'Brien, K. 2003, ApJ, 583, L95
\smallskip

Steeghs, D. \& Casares, J., 2002, ApJ, 568, 273
\smallskip

O'Brien, K. 2001 in ASP Conf. Series, (astro-ph/0110267)
\smallskip

Vrtilek, S. D., Raymond, J. C., Garcia, M. R., Verbunt, F., Hasinger, G., \& Kurster, M. 1990,  A$\&$A, 235, 162}

\end{document}